\documentclass[twocolumn]{jpsj3} %% two-column layout
\title{Pressure Dependence of Superconducting Transition Temperature on Perovskite-Type Fe-Based Superconductors and NMR Study of Sr$_2$VFeAsO$_3$}

\author{Hisashi \textsc{Kotegawa}$^{1,4}$\thanks{E-mail address: kotegawa@crystal.kobe-u.ac.jp}, Yuuki \textsc{Tao}$^{1}$, Hideki \textsc{Tou}$^{1,4}$, Hiraku \textsc{Ogino}$^{3,4}$, Sigeru \textsc{Horii}$^{5}$, Kohji \textsc{Kishio}$^{3,4}$, and Jun-ichi \textsc{Shimoyama}$^{3,4}$}

\inst{$^{1}$Department of Physics, Kobe University, Kobe 657-8501\\
$^{2}$Division of Molecular Materials Science, Graduate School of Science, Osaka City University, Osaka 558-8585\\
$^{3}$Department of Applied Chemistry, The University of Tokyo, Tokyo 113-8656\\
$^{4}$JST, Transformative Research - Project on Iron Pnictides (TRIP), Chiyoda, Tokyo 102-0075\\
}

\abst{
We report the pressure dependences of the superconducting transition temperature ($T_c$) in several perovskite-type Fe-based superconductors through the resistivity measurements up to $\sim4$ GPa.
In Ca$_4$(Mg,Ti)$_3$Fe$_2$As$_2$O$_{y}$ with the highest $T_c$ of 47 K in the present study, the $T_c$ keeps almost constant up to $\sim1$ GPa, and starts to decrease above it.
From the comparison among several systems, we obtained a tendency that low $T_c$ with the longer $a$-axis length at ambient pressure increases under pressure, but high $T_c$ with the shorter $a$-axis length at ambient pressure hardly increases.
We also report the $^{75}$As-NMR results on Sr$_2$VFeAsO$_3$.
NMR spectrum suggests that the magnetic ordering occurs at low temperatures accompanied by some inhomogeneity.
In the superconducting state, we confirmed the anomaly by the occurrence of superconductivity in the nuclear spin lattice relaxation rate $1/T_1$, but the spin fluctuations unrelated with the superconductivity are dominant.
It is conjectured that the localized V-$3d$ moments are magnetically ordered and their electrons do not contribute largely to the Fermi surface and the superconductivity in Sr$_2$VFeAsO$_3$.

}

\kword{Perovskite-type Fe-based superconductor, Sr$_2$VFeAsO$_3$, pressure effect, NMR}

\begin{document}
\maketitle

\section{Introduction}

\begin{figure}[b]
\centering
\includegraphics[width=0.9\linewidth]{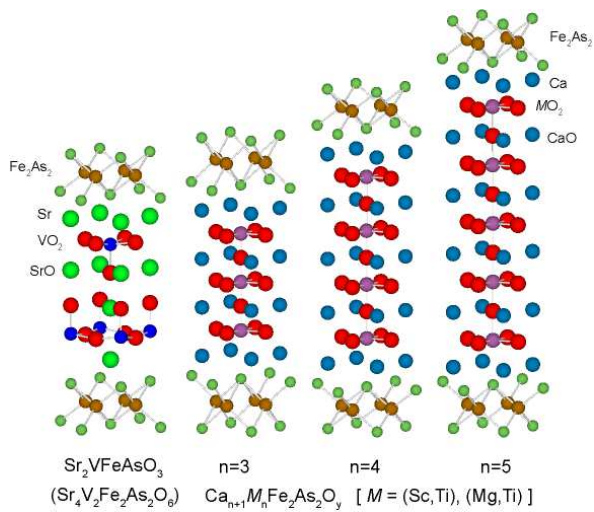}
\caption[]{(color online) The crystal structures of Sr$_2$VFeAsO$_3$ and Ca$_{n+1}$$M_n$Fe$_2$As$_2$O$_{y}$ with $n=3, 4, 5$ [$y \sim 3n-1$, $M=$ (Sc,Ti) and  (Mg,Ti)]. 
}
\end{figure}

In Fe-based superconductors, two dimensional layers composed of Fe and pnictogen (or chalcogen) are stacked with sandwiching the block layers or ions.
FeSe is a binary compound without block layers,\cite{Hsu} and 111 systems such as LiFeAs or 122 systems such as (Ba,K)Fe$_2$As$_2$ have ions between FeAs layers.\cite{Wang,Rotter}
Firstly discovered 1111 systems possess relatively thick block layers,\cite{Kamihara} and afterward 21113 systems such as Sr$_2$ScFePO$_3$ and Sr$_2$VFeAsO$_3$ with thick perovskite-layers were discovered.\cite{Ogino,Zhu}
Characteristic point of Fe-based superconductors is richness of the structural kinds.
Recently, in the perovskite-type Fe-based superconductors, Ogino and co-workers discovered some new superconductors with the relatively high superconducting transition temperature ($T_c$) of 40 K class.\cite{Ogino2,Ogino3,Kawaguchi,Shimizu,Ogino4}
For example, their compositions and $T_c^{onset}$ are Ca$_4$(Sc$_{0.67}$Ti$_{0.33}$)$_3$Fe$_2$As$_2$O$_y$ ($T_c^{onset} = 33$ K), Ca$_5$(Sc$_{0.5}$Ti$_{0.5}$)$_4$Fe$_2$As$_2$O$_y$ ($T_c^{onset} = 41$ K), Ca$_6$(Sc$_{0.4}$Ti$_{0.6}$)$_5$Fe$_2$As$_2$O$_y$ ($T_c^{onset} = 42$ K), and Ca$_4$(Mg$_{0.25}$Ti$_{0.75}$)$_3$Fe$_2$As$_2$O$_{y}$ ($T_c^{onset} = 47$ K).
Here $T_c^{onset}$ indicates the onset temperature of resistivity drop.
Figure 1 shows the crystal structures of these compounds indicated Ca$_{n+1}$$M_n$Fe$_2$As$_2$O$_{y}$ [$n=3, 4, 5$, $y \sim 3n-1$, $M=$ (Sc,Ti) and (Mg,Ti)] and Sr$_2$VFeAsO$_3$.
The $T_c$ of more than 40 K exceeding the maximum $T_c$ of 11, 111, and 122 systems implies that two-dimensional crystal structure is favorable for high $T_c$.
These findings give us an opportunity to investigate an importance of two dimensionality of crystal structure for high $T_c$, but first we need to know whether these $T_c$'s are indeed the maximum of the systems.
We have two effective ways to control $T_c$ in Fe-based superconductors.
One is a chemical carrier doping, but unfortunately it is difficult to obtain the sample of the single phase with different composition from the present sample.
Another way to control $T_c$ is applying pressure.
For instance, the material showing the most drastic change under pressure is FeSe.
Its $T_c$ of 8 K increases to 37 K under pressure.\cite{Mizuguchi,Margadonna2,Medvedev,Masaki}
From comparisons of the structures and $T_c$ in FeSe under pressure and in some Fe-based superconductors, the height of pnictogen or chalcogen from Fe-plane ($h$) is suggested to be the most important factor for high $T_c$.\cite{Mizuguchi2,Okabe,Kotegawa} 
On the other hand, we have investigated the pressure dependence of $T_c$ in Sr$_2$ScFePO$_3$ and Sr$_2$VFeAsO$_3$.\cite{Kotegawa}
The $T_c$ of 17 K in Sr$_2$ScFePO$_3$ with lower $h$ decreases drastically, while the $T_c$ of 37 K in Sr$_2$VFeAsO$_3$ with higher $h$ increases up to 46 K under pressure.\cite{Kotegawa}
If we recognize that the $h$ has the optimized value between those of two compounds and that $h$ decreases under pressure in both compounds, this contrasting pressure effect can be explained qualitatively.
Thus we consider that pressure is an effective way to control $T_c$ through a change of $h$ as a main factor.
Although $h$ is not determined experimentally in most of new perovskite-type superconductors at present, Sr$_2$VFeAsO$_3$ with similar structure shows the drastic increase in $T_c$.
The pressure effects for the new superconductors are promising for obtaining higher $T_c$.

\begin{figure}[b]
\centering
\includegraphics[width=0.75\linewidth]{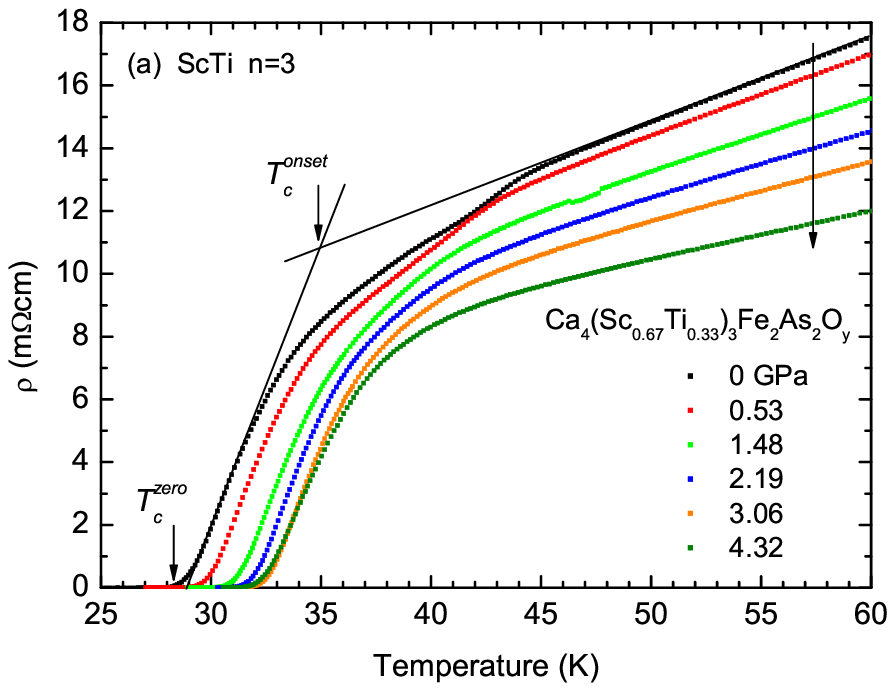}
\includegraphics[width=0.75\linewidth]{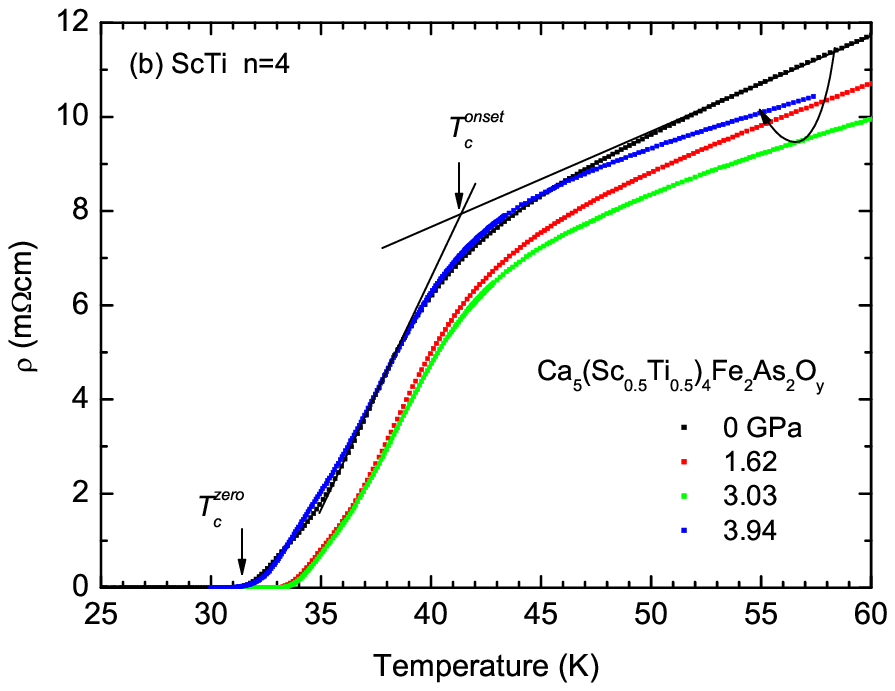}
\includegraphics[width=0.75\linewidth]{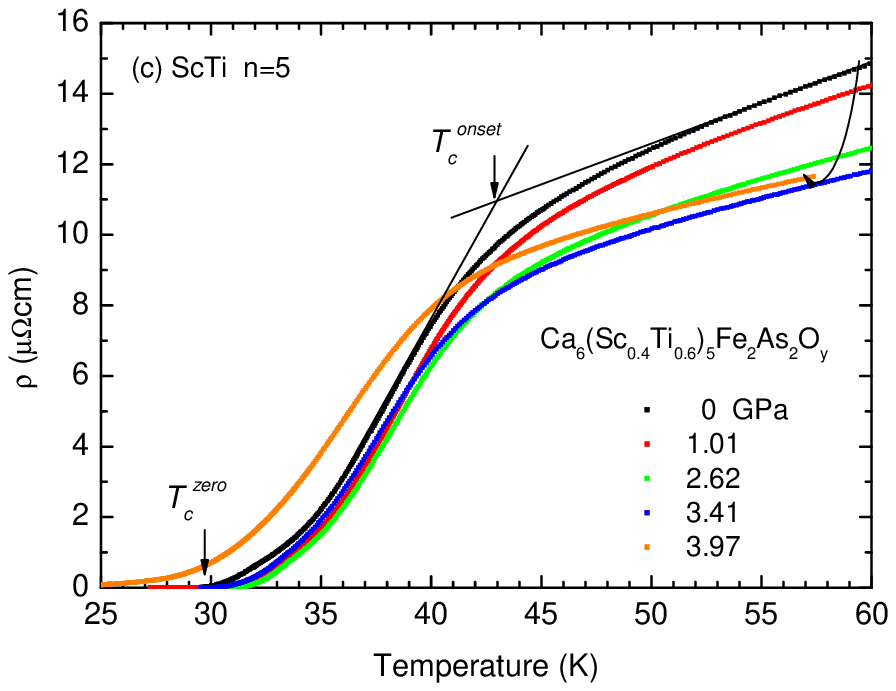}
\includegraphics[width=0.75\linewidth]{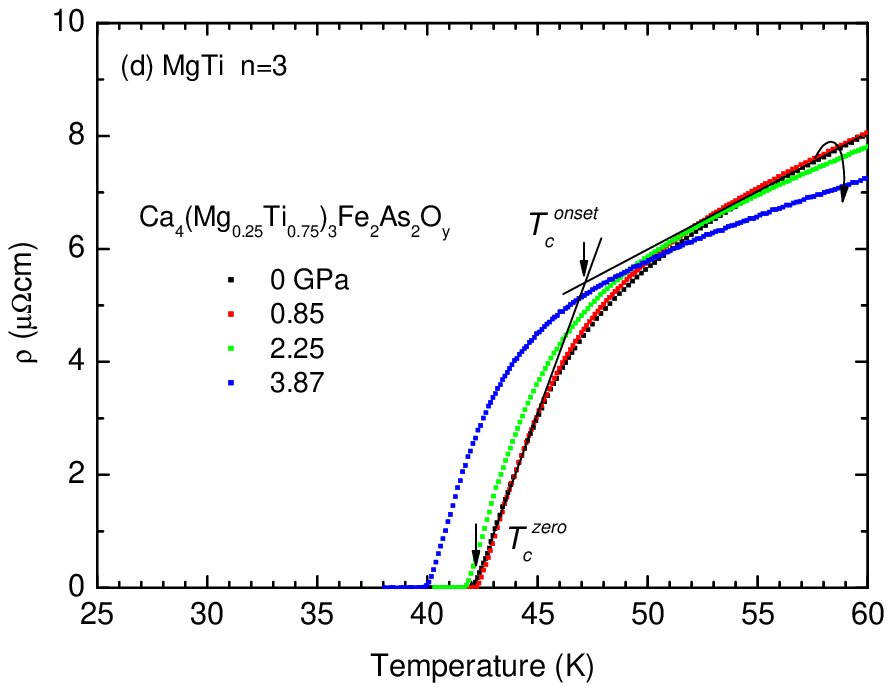}
\caption[]{(color online) Temperature dependences of resistivity under pressures up to $\sim4$ GPa in Ca$_{n+1}$(Sc,Ti)$_n$Fe$_2$As$_2$O$_{y}$ ($n=3, 4, 5$, $y \sim 3n-1$) and Ca$_4$(Mg,Ti)$_3$Fe$_2$As$_2$O$_{y}$. The $T_c^{onset}$ and $T_c^{zero}$ are defined as shown in the figure. 
}
\end{figure}

On the other hand, Sr$_2$VFeAsO$_3$ has a different issue.
The band calculation has suggested the V orbitals contribute to the formation of the Fermi surface.\cite{Shein1}
The existence of the same nesting properties as those of other Fe-based superconductors, which are believed to be involved with superconductivity, seems to be controversial in Sr$_2$VFeAsO$_3$.\cite{KWLee,Mazin,Nakamura}
It should be clarified whether the superconductivity in Sr$_2$VFeAsO$_3$ is understood in the same framework as those in other Fe-based superconductors.
Quite recently Qian {\it et al.} reported that the similar nesting properties to other Fe-based superconductors remain in this compounds through an angle-resolved photoemission spectroscopy.\cite{Qian}
In this paper, we performed $^{75}$As-NMR measurement on Sr$_2$VFeAsO$_3$ and the resistivity measurements under pressure in some 21113 systems.

\section{Experimental Procedure}

Polycrystalline samples were synthesized by solid-state reaction.\cite{Ogino2,Ogino3}
Electrical resistivity ($\rho$) measurement at high pressures up to $\sim4$ GPa was carried out using an indenter cell.\cite{indenter}
Electrical resistivity was measured by a four-probe method using silver paste for contact.
The typical sample dimensions were $0.8\times0.3\times0.2$ mm$^3$.
We used Daphne 7474 for the resistivity measurement as a pressure-transmitting medium.\cite{Murata}
Applied pressure was estimated from the $T_{c}$ of the lead manometer.
The NMR measurement was performed by a standard spin-echo method.
The powdered sample are partially oriented under magnetic field by a mechanical shock.

\section{Results and Discussion}

\subsection{Pressure effect in Perovskite-type Fe-based superconductors }

\begin{figure}[t]
\centering
\includegraphics[width=0.75\linewidth]{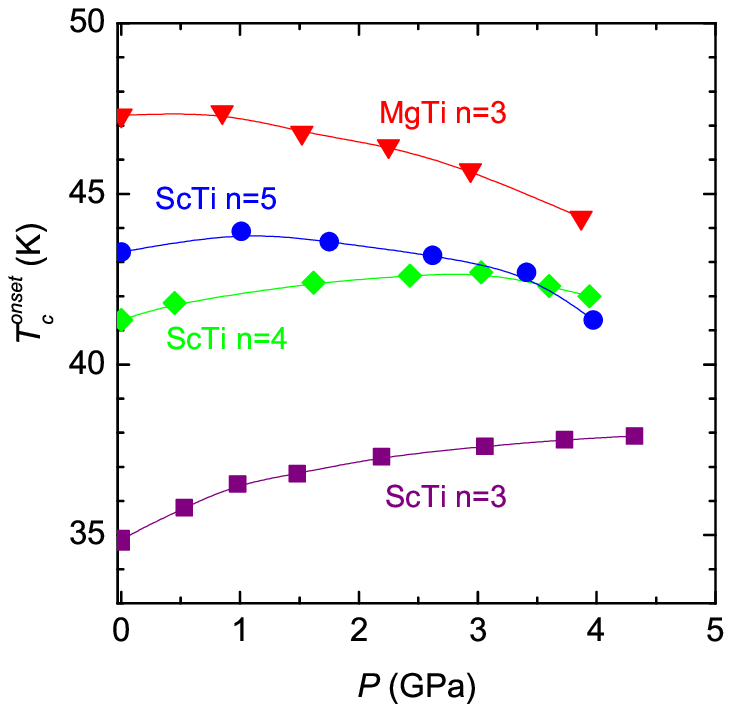}
\includegraphics[width=0.75\linewidth]{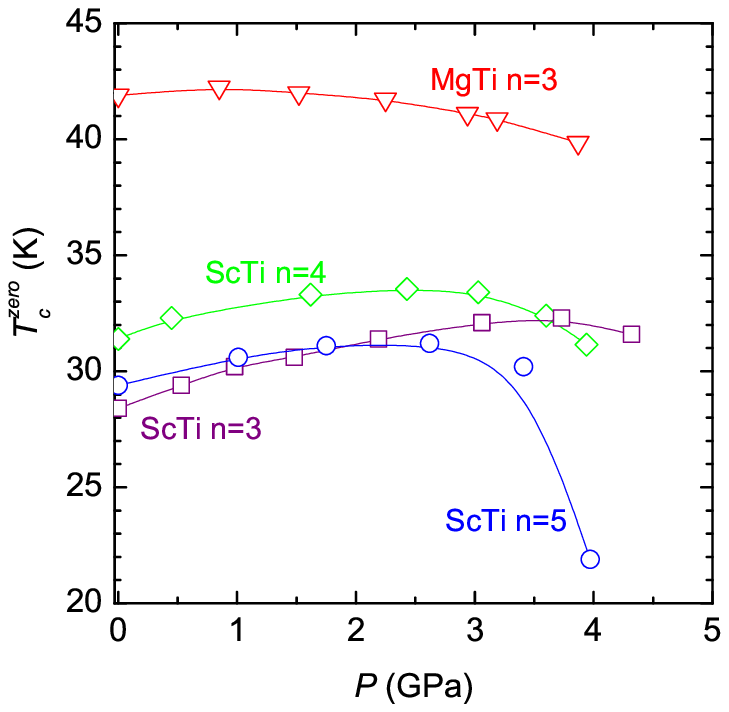}
\caption[]{(color online) Pressure dependences of $T_c^{onset}$ and $T_c^{zero}$ in Ca$_{n+1}$(Sc,Ti)$_n$Fe$_2$As$_2$O$_{y}$ ($n=3, 4, 5$, $y \sim 3n-1$) and Ca$_4$(Mg,Ti)$_3$Fe$_2$As$_2$O$_{y}$.
}
\end{figure}

Figure 2 shows the temperature dependences around $T_c$ in Ca$_{n+1}$(Sc,Ti)$_n$Fe$_2$As$_2$O$_{y}$ ((a): ScTi $n=3$, (b): ScTi $n=4$, and (c): ScTi $n=5$) and Ca$_4$(Mg,Ti)$_3$Fe$_2$As$_2$O$_{y}$ ((d): MgTi $n=3$).
Each $T_c^{onset}$ and $T_c^{zero}$ is determined as shown at ambient pressure in the figure.
In the ScTi system with $n=3$, $T_c$ increases under pressure and almost saturates at around 4 GPa. The increase in $T_c^{onset}$ is estimated to be 3.1 K at 4.32 GPa.
Similarly the ScTi system with $n=4$ also shows an increase in $T_c$ under pressure, but $T_c$'s are almost same between 1.62 and 3.03 GPa.
The $T_c$ obviously decreases at 3.94 GPa.
In the case of $n=5$, an increase in $T_c$ is less than 1 K, and $T_c$ decreases above 3 GPa.
In the MgTi system with $n=3$ which possesses the maximum $T_c$ among these compounds, $T_c$ is almost unchanged between 0 and 0.85 GPa, and it starts to decrease above $\sim1$ GPa.

\begin{figure}[b]
\centering
\includegraphics[width=0.75\linewidth]{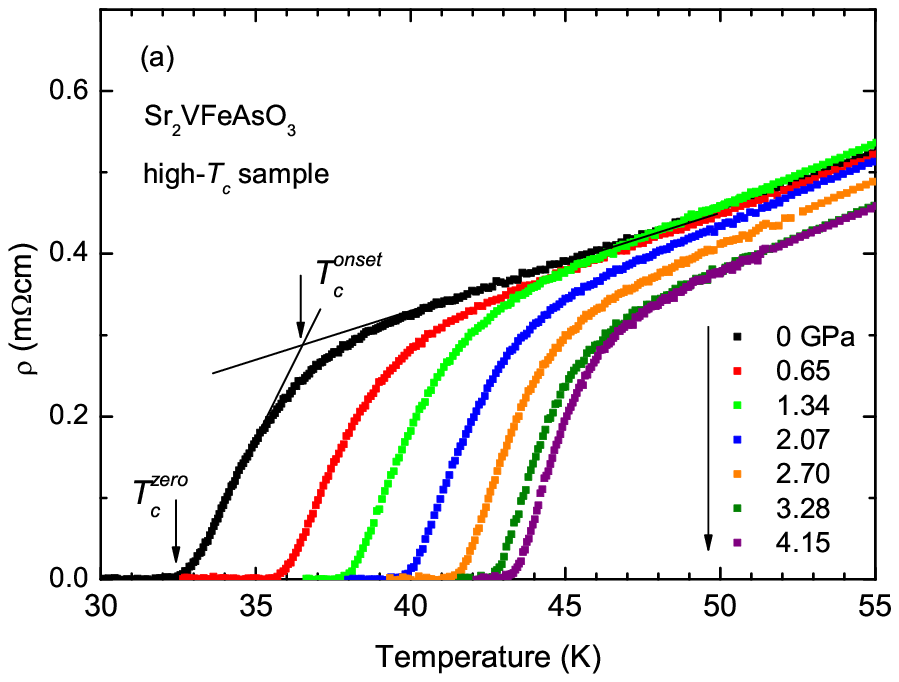}
\includegraphics[width=0.75\linewidth]{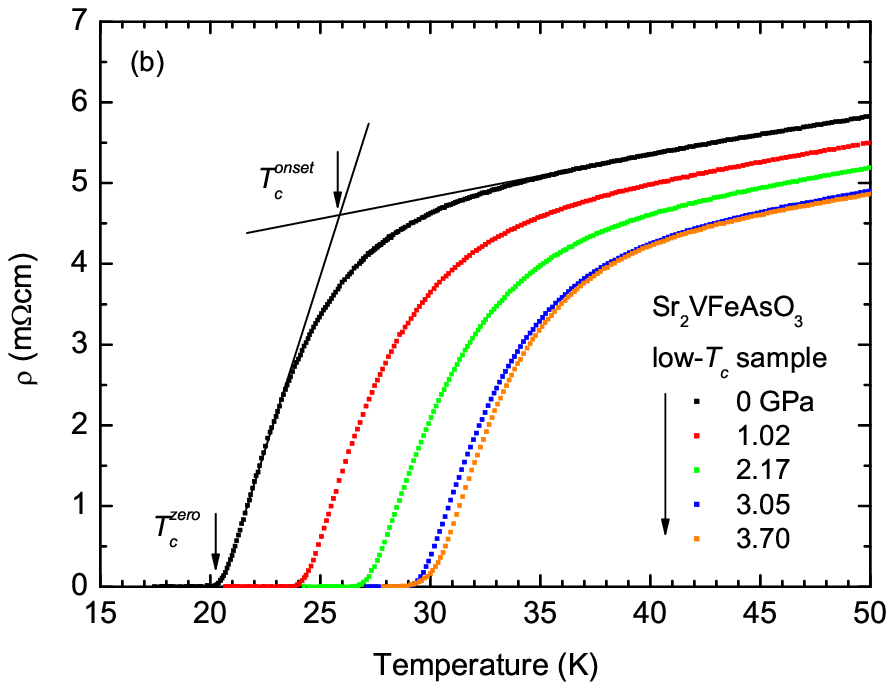}
\includegraphics[width=0.75\linewidth]{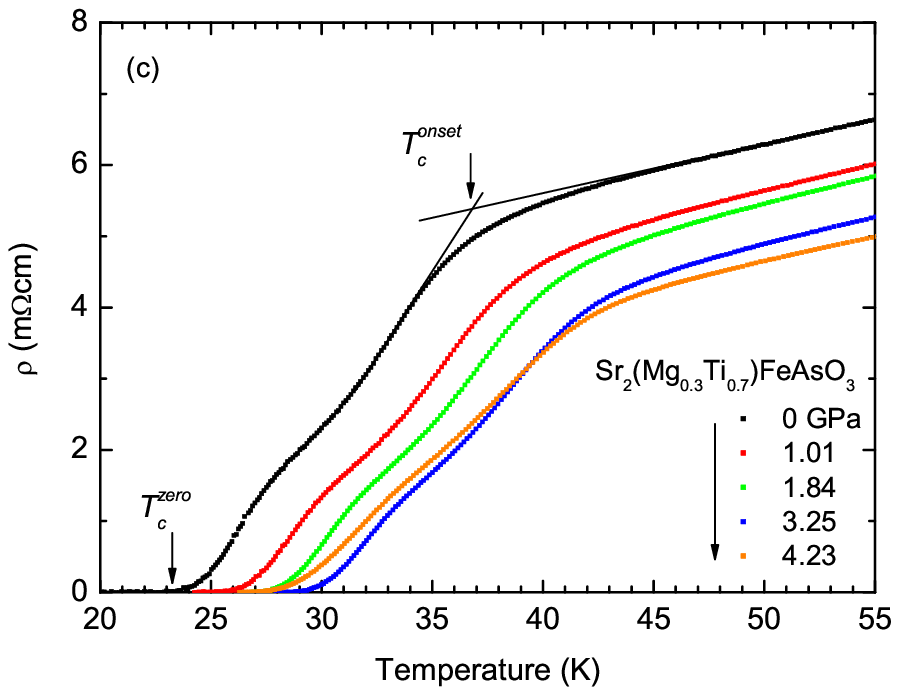}
\includegraphics[width=0.75\linewidth]{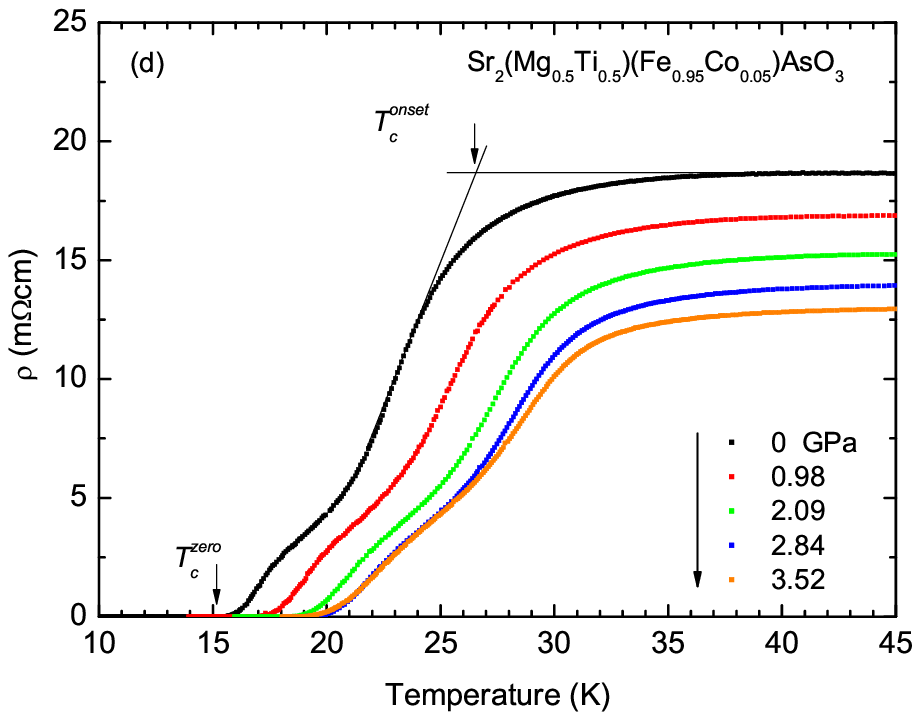}
\caption[]{(color online) Temperature dependences of resistivity under pressures up to $\sim4$ GPa in Sr$_2$VFeAsO$_3$ with high $T_c$ and low $T_c$, Sr$_2$(Mg$_{0.3}$Ti$_{0.7}$)FeAsO$_3$, and Sr$_2$(Mg$_{0.5}$Ti$_{0.5}$)(Fe$_{0.95}$Co$_{0.05}$)AsO$_3$.
}
\end{figure}

Figure 3 shows the pressure dependences of $T_c^{onset}$ and $T_c^{zero}$.
We can see a clear tendency that $T_c$ which is lower at ambient pressure can increase under pressure but $T_c$ which is already high at ambient pressure does not increase any more.
This tendency is reminiscent of that in 1111 systems,\cite{Takahashi,Takeshita} and seems to indicate that the $T_c$ of around 47 K is the maximum in these perovskite-type systems.
At least, applying pressure is not the effective way to obtain much higher $T_c$ in these systems.

Next we report pressure effect on 21113 systems.
We have already investigated the pressure effect in Sr$_2$VFeAsO$_3$.\cite{Kotegawa}
Figure 4(a) is the same data as the previous report.
$T_c^{onset}$ of 36.4 K increased up to 46.0 K under $\sim4$ GPa.
Figure 4(b) shows the results in Sr$_2$VFeAsO$_3$ with lower $T_c$.
Han {\it et al.} synthesized the sample with different $T_c$ with changing the content of oxygen in the nominal composition.\cite{Han}
At present we cannot conclude what is difference between high-$T_c$ sample and low-$T_c$ sample from structural analysis, but it likely originates from the difference in carrier density on the analogy of their Hall effect measurement.\cite{Han}
As shown in the figure, this compound also shows the large increase of $\sim 10$ K in $T_c$ under pressure.
Figures 4(c) and (d) show the pressure effects on Sr$_2$(Mg$_{0.3}$Ti$_{0.7}$)FeAsO$_3$ and Sr$_2$(Mg$_{0.5}$Ti$_{0.5}$)(Fe$_{0.95}$Co$_{0.05}$)AsO$_3$.\cite{Sato}
The superconductivity in each compound is induced by the different method of electron doping and the $T_c$'s in two compounds are different.
However, the $T_c$'s in both compounds increase similarly under pressure and almost saturate at around 4 GPa.

\begin{figure}[htb]
\centering
\includegraphics[width=0.75\linewidth]{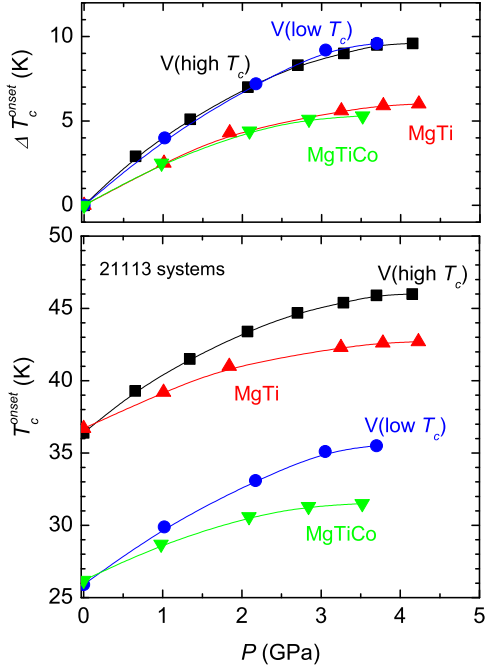}
\caption[]{(color online) Pressure dependences of $T_c^{onset}$ and $\Delta T_c^{onset}$ in Sr$_2$VFeAsO$_3$ with high $T_c$ and low $T_c$, Sr$_2$(Mg$_{0.3}$Ti$_{0.7}$)FeAsO$_3$, and Sr$_2$(Mg$_{0.5}$Ti$_{0.5}$)(Fe$_{0.95}$Co$_{0.05}$)AsO$_3$.
}
\end{figure}

Figure 5 shows the pressure dependence of $T_c^{onset}$ in four 21113 systems.
All the compounds shows the increase in $T_c$ and seems to saturate at around 4 GPa.
The top panel shows the pressure dependence of $\Delta T_c^{onset}$ which is difference in $T_c^{onset}$ between at ambient pressure and under pressure.
We can see the clear overlap in each compound for V systems and for MgTi systems, even though the original $T_c$'s are different.
Note that we cannot observe such an overlap in pressure dependence of $T_c^{onset}(P) / T_c^{onset}(P=0 {\rm GPa})$.
The difference in $T_c$ at ambient pressure is considered to attribute to mainly the difference in the carrier density.
The clear overlap implies that the increase in $T_c$ under pressure can be explained without considering the change in the carrier density, that is, the pressure variation of the crystal structure is an important factor leading to the increase in $T_c$.
In addition, the similar pressure curves in $P$ vs. $T_c^{onset}$ under pressure between V systems and MgTi systems impresses that the superconductivity in both compounds occur in the same framework, that is, the electrons in the block layer are unrelated with the superconducting mechanism.

Table 1 indicates the $T_c^{onset}$ at ambient pressure,  the maximum $\Delta T_c^{onset}$ ($\Delta T_{c,max}^{onset}$), and the pressure with the maximum $\Delta T_c^{onset}$ ($P_{max}$), the $a$-axis length, and the pnictogen-height, $h$.
The $T_c^{onset}$'s in the present samples do not exceed 50 K even under pressure.
Rather applying pressure is ineffective for raising $T_c$ for the material which possesses already high $T_c$.
As already pointed out, these systems have a tendency that $T_c$ at ambient pressure increases as the $a$-axis length decreases.\cite{Ogino3}
The relationship between the $a$-axis length and $h$ is not clear in these systems, but if we recognize that $h \sim 1.38$ \AA~ in NdFeAsO$_{1-y}$ is the optimized value,\cite{Mizuguchi2,Okabe} the MgTi system with $n=3$ might have $h$ close to $1.38$ \AA.
Otherwise, $h$'s of these compounds might be higher $h$ than $1.38$ \AA~ at ambient pressure and have a tendency to decrease largely under pressure as the $a$-axis length is longer.
Anyway, detailed structural analyses are desired for understanding what governs the $T_c$ in these systems.

\begin{table}[ht]

 \begin{center}
  \begin{tabular}{lccccc}
       & $T_c^{onset}$ & $\Delta T_{c,max}^{onset}$  &  $P_{max}$ &  $a$   & $h$  \\
   \hline
          &    &   &                  &       &  \\
MgTi21113 & 36.7 &  6.0 &  $\ge4$    &  3.936  & $\ast$  \\
V21113    & 36.4 &  9.6 &   $\ge4$   & 3.930  & 1.42  \\
ScTi n=3  & 34.8 & 3.1  & $\ge4$  & 3.922  & - \\
ScTi n=4  & 41.3 & 1.4 & 3   & 3.902  & -  \\
ScTi n=5  & 43.3 & 0.6 & 1   &  3.884 & - \\
MgTi n=3  & 47.3 & $\sim$0.1 & $\sim$0.5 & 3.877  & - \\

                               &    &         &           &    &     \\ 
   \hline
  \end{tabular}
    \caption{$T_c^{onset}$ at ambient pressure (K),  the maximum $\Delta T_c^{onset}$ (K), and the pressure with the maximum $\Delta T_c^{onset}$ (GPa), $a$-axis length (\AA), and $h$ (\AA). The $h$ indicates the height of As from Fe-plane. The structural data were obtained from other paper.\cite{Sato,Zhu,Ogino2,Ogino3} The $h$ of some compounds indicated by - are still unclear. ($\ast$):In MgTi21113, we measured the pressure effect on SrMg$_{0.3}$Ti$_{0.7}$FeAsO$_3$, and $h$ is estimated to be 1.40 \AA~ for SrMg$_{0.5}$Ti$_{0.5}$FeAsO$_3$.\cite{Ogino5}
    }
     \end{center}
\end{table}

\subsection{$^{75}$As-NMR Study in Sr$_2$VFeAsO$_3$}

\begin{figure}[b]
\centering
\includegraphics[width=0.8\linewidth]{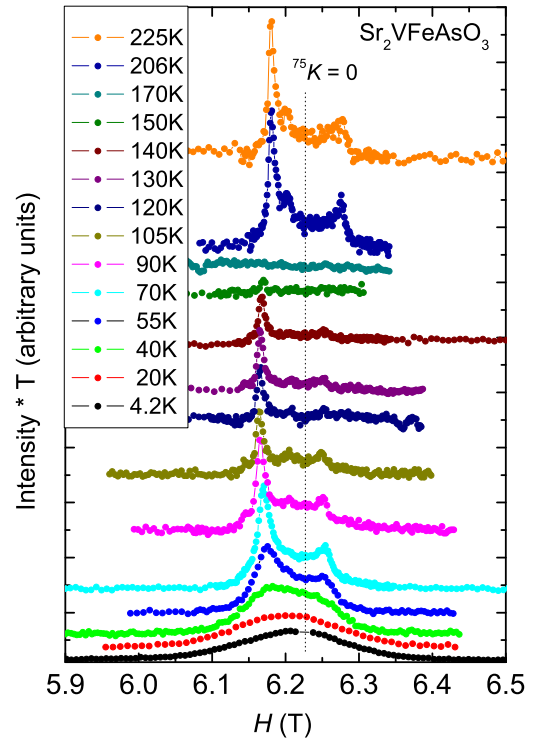}
\caption[]{(color online) $^{75}$As-NMR spectra at several temperatures. The signal disappears at $155-190$ K. The spectrum markedly broadens at low temperatures, indicative of the appearance of the internal field.
}
\end{figure}

We demonstrate $^{75}$As-NMR measurement in Sr$_2$VFeAsO$_3$ to clarify the V-$3d$ electronic state and the properties of the superconductivity.
Figure 6 shows $^{75}$As-NMR spectra for the transition of $-1/2 \leftrightarrow 1/2$ measured at $\sim6.2$ T and at several temperatures.
At high temperatures, the spectrum exhibits the powder pattern affected by quadrupolar interaction.
The quadrupole frequencies are estimated to be 7.23 MHz at 4.2 K and 7.77MHz at 210 K from the splitting of $-1/2 \leftrightarrow -3/2$ and $1/2 \leftrightarrow 3/2$ transitions (not shown).
The distinct peak observed at lower field corresponds to $H \parallel ab$, meaning that the sample is partially oriented to $H \parallel ab$.
In the temperature range of $155-190$ K, the signal completely disappears owing to the fast relaxation, indicating that some phase transition occurs at $\sim170$ K.
Note that this temperature range of the signal loss is unchanged between 6.2 T and 12 T.
The signal recovers below 150 K but it is still small.
At low temperature, especially below 70 K, the spectrum obviously broadens and the characteristic shape of the powder pattern is lost below 40 K.
This broadening strongly suggests that the internal field appears at low temperature in this system, because we confirmed the similar broadening independently of applied magnetic field.

\begin{figure}[htb]
\centering
\includegraphics[width=0.8\linewidth]{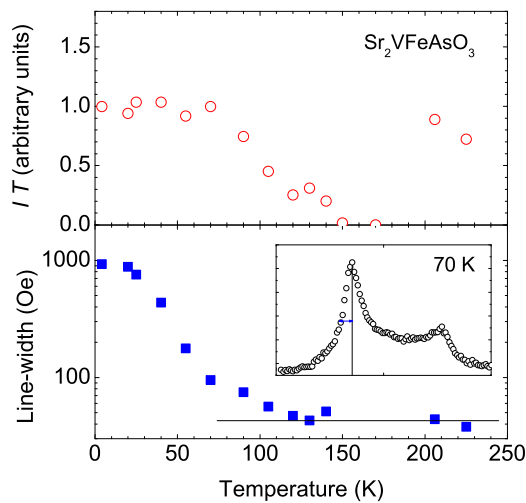}
\caption[]{(color online) The temperature dependences of the intensity $\times$ temperature ($I T$), and line-width. The arrow of the inset corresponds to the estimated line-width.
}
\end{figure}

Figure 7 shows the temperature dependences of the product of the intensity and  temperature ($I T$), and line-width.
Here the intensity was obtained by integrating all the spectrum of the powder pattern without a correlation for nuclear spin-spin relaxation time, and the line-width was obtained from half-width as shown in the inset.
$I T$ should keep constant in principle, and certainly shows constant below 70 K.
It does not recover in the temperature range of $90-150$ K, meaning that we do not observe the whole part in the sample in this temperature range.
The line-width of $120-150$ K is still as narrow as that above 200 K, indicating that the internal field at As site is quite small or zero in the region where we observe through NMR.
The line-width increases moderately from $\sim 120$ K down to 70 K accompanied by the recovery of the signal intensity.
We can conjecture that the internal field is already induced even above 120 K in the region where we cannot observe owing to the fast relaxation.
On the other hand, the line-width is rapidly increases below $\sim70$ K.
The observation of the clear powder pattern at 70 K, where the signal is almost recovered, indicates that large internal field does not appear along any directions at the As site in this temperature.
The rapid increase in the line-width suggests the rearrangement of the magnetic structure or the development of the ordered moment.

The spectrum at 4.2 K is symmetric with the center at the position close to $K=0$.
If the internal field at the As-site lies antiferromagnetically in the $ab$ plane, the partially oriented powder pattern along $H//ab$ is expected to change to the rectangular shape with the sharp edge, but this does not match the present spectrum.
The internal field is rather suggested to turn along the $c$ axis at the As-site, although the large broadening of the spectrum seems to exclude the perfect alignment along the c-axis.
Recently, Cao {\it et al.} suggested that the ordered moment is absent at the Fe site, and Tegel {\it et al.} showed the possible incommensurate magnetic ordering of V moment with a propagation vector of $q=(0,0,0.306)$.\cite{Cao,Tegel}
Unfortunately, it is difficult to address the magnetic structure form the present NMR spectrum, because we cannot obtain the hyperfine coupling constants at present.
However our experiments strongly suggest that the existence of the magnetic ordering and the change in the magnetic state at $\sim 70$ K.
Cao {\it et al.} also observed two magnetic transitions at $\sim150$ K and $\sim55$ K.\cite{Cao}
The increase in the internal field at the As site below 70 K most likely corresponds to the second phase transition.

\begin{figure}[htb]
\centering
\includegraphics[width=0.75\linewidth]{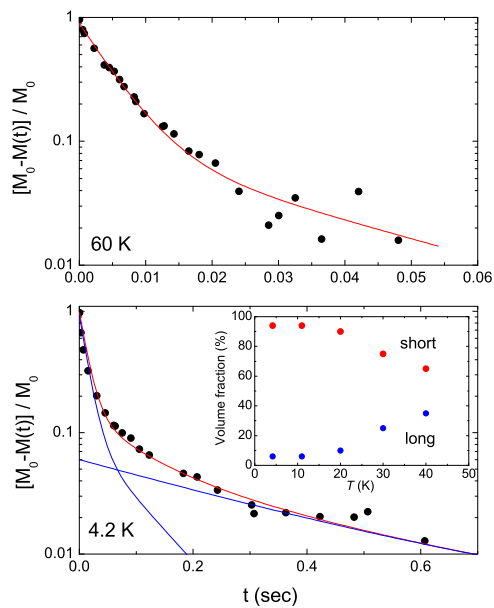}
\caption[]{(color online) The recovery curves at 60 K and 4.2 K. The $T_1$ is uniquely determined at 60 K, but we used two-components fitting below 40 K. The inset shows the temperature dependence of each volume fractions. The short component is dominant at low temperatures.
}
\end{figure}

Figure 8 shows the recovery curves to obtain the nuclear spin-lattice relaxation time $T_1$.
The red curves are the fitting results, which follow the experimental data well at 60 K.
However, we cannot fit the data by one component below $\sim40$ K, and we used two-components fitting to obtain the short component and the long component in $T_1$.
As shown in the figure, this two-components fitting follows the experimental data well, but this indicates that the electronic state is spatially inhomogeneous.
The inset shows the temperature dependence of the volume fraction of both components.
At low temperatures, it is found that the short component is dominant in the sample.

\begin{figure}[htb]
\centering
\includegraphics[width=0.8\linewidth]{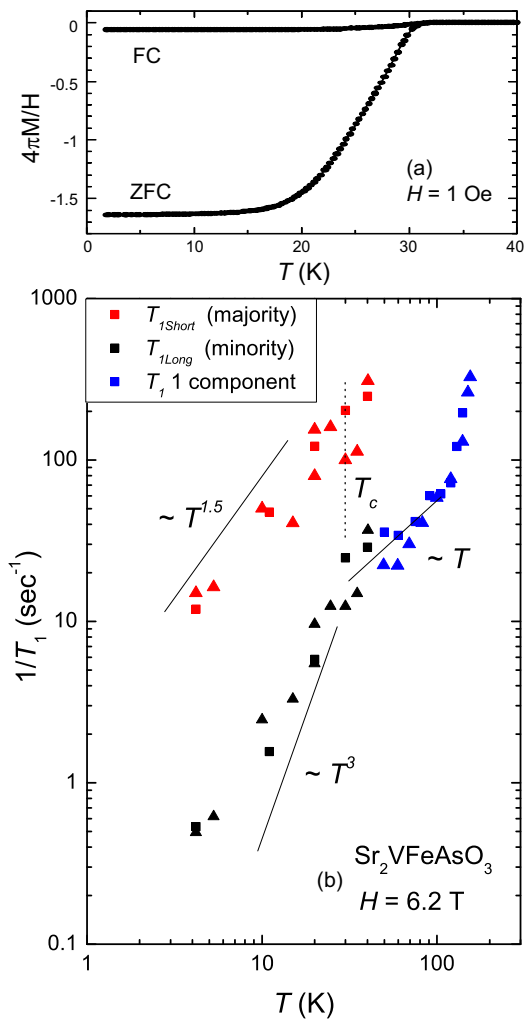}
\caption[]{(color online) Temperature dependence of magnetization and $1/T_1$ in Sr$_2$VFeAsO$_3$. The sample shows a clear diamagnetic signal. FC and ZFC indicate field cooling and zero-field cooling, respectively. $T_1$ is not determined uniquely below 40 K corresponding to the inhomogeneous of the electronic state. We observed the significant decrease in $1/T_1$ in the long component below $T_c$, but the majority part at low temperatures is the short component which is insensitive to the occurrence of the superconductivity. We measured two different samples which are distinguished by the squares and the triangles.
}
\end{figure}

Figure 9 shows the temperature dependence of magnetization and $1/T_1$.
We checked two different samples and obtained almost same data.
The $1/T_1$ shows the rapid increase toward $\sim 170$ K, indicative of the critical slowing down of the spin fluctuations at the phase transition of $\sim170$ K.
The relation roughly close to $T_1T =$ constant is observed below 120 K, but the short component in $T_1$ ($1/T_{1Short}$) rapidly appears below 40 K accompanied by the broadening of the spectrum.
The long component in $1/T_1$ ($1/T_{1Long}$) is smoothly connected to the $1/T_1$ of 1 component.
It also shows the decrease below $T_c$ with the slop slightly slower than $T^3$.
However, $1/T_{1Short}$, which is majority at low temperatures, seems not to be affected by the occurrence of the superconductivity, although the sample shows a clear Meissner effect of more than 100 \% in the volume fraction.
This behavior resembles that of PrFeAsO$_{1-y}$ system where the spin fluctuations from Pr-localized moments are dominant compared with that from the conduction electrons.\cite{Yamashita}
This indicates that the spin fluctuations of some localized moment, presumably V moment, are dominant below 40 K in Sr$_2$VFeAsO$_3$, because the Fe 3d electrons contribute to the main part of the Fermi surface.\cite{Qian}
The slow temperature dependence of $1/T_{1Short}$ below $T_c$ suggests that there is no large superconducting gap in the V electronic state.
The $1/T_{1Long}$ is speculated to originate from the region where the spin fluctuations of the localized moment are accidentally canceled or absent owing to the inhomogeneity in the magnetic structure or the valence of Vanadium.

\subsection{Summary}

We have investigated the pressure dependence of $T_c$ in some perovskite-type Fe-based superconductors through the resistivity measurements.
We observed the tendency that the $T_c$ does not increase in the sample whose $T_c$ is already high at ambient pressure.
For example, the highest $T_c^{onset} = 47$ K in this study for Ca$_4$(Mg,Ti)$_3$Fe$_2$As$_2$O$_{y}$ does not increase under pressure.
In 21113 system, we observed the good scaling in the pressure dependence of the increase in $T_c$ for each compound irrespective of the difference of the carrier density.

We also performed NMR measurement on Sr$_2$VFeAsO$_3$.
We found that the phase transition with the strong enhancement of $1/T_1$ occurs at $\sim170$ K.
The internal field at As site develops rapidly below $\sim70$ K.
The $T_1$ becomes inhomogeneous below 40 K accompanied by the development of the internal field and the spin fluctuations giving the short $T_1$ become dominant at low temperatures.
This major spin fluctuation does not show an obvious anomaly for the occurrence of the superconductivity.
We can conjecture that the V-$3d$ magnetic moments are almost localized and ordered in this system.
Even if V-$3d$ electrons somewhat contribute to the Fermi surface, the superconducting gap in the V orbitals is expected to be quite small.

\subsection*{Acknowledgement}

We acknowledge Mr. S. Sato and Y. Shimizu of The University of Tokyo
for cooperation of sample preparation.
This work has been partly supported by Grants-in-Aid for Scientific Research (Nos. 19105006, 20740197, 20102005, 22013011, and 22710231) from the Ministry of Education, Culture, Sports, Science, and Technology (MEXT) of Japan.

\end{document}